\documentclass[conference]{IEEEtran}
\usepackage{color}
\usepackage[utf8]{inputenc}
\usepackage{amsmath}

\usepackage{amssymb}
\usepackage[hyphens]{url}
\usepackage[ruled, linesnumbered]{algorithm2e}
\usepackage[pdftex,outerbars]{changebar}
\nochangebars

\usepackage{graphicx}

\usepackage[disable]{todonotes}

\usepackage{enumitem}
\usepackage{upgreek}
\newcommand{\q}[1]{``#1''}

\title{Distributing and Obfuscating Firewalls via Oblivious Bloom Filter Evaluation} 
\author{\IEEEauthorblockN{Ken Goss}
	\IEEEauthorblockA{
		University of Missouri\\
		Columbia, Missouri\\
		Email: kgoss@mail.missouri.edu}
	\and
	\IEEEauthorblockN{Wei Jiang}
	\IEEEauthorblockA{
		University of Missouri\\
		Columbia, MO\\
		Email: wjiang@missouri.edu}
}

\newcommand{\compBogThree}{$27\ell$}

\begin{document}
\maketitle
\begin{abstract}
Firewalls have long been in use to protect local networks from threats of the larger Internet. 	Although firewalls are effective in preventing attacks initiated from outside, they are vulnerable to insider threats, e.g., malicious insiders may access and alter firewall configurations, and disable firewall services. In this paper, we develop an innovative distributed architecture to obliviously manage and evaluate firewalls to prevent both insider and external attacks oriented to the firewalls. Our proposed structure alleviates these issues by obfuscating the firewall rules or policies themselves, then distributing the function of evaluating these rules across multiple servers. Thus, both accessing and altering the rules are considerably more difficult thereby providing better protection to the local network as well as greater security for the firewall itself. We achieve this by integrating multiple areas of research such as secret sharing schemes and multi-party computation, as well as Bloom filters and Byzantine agreement protocols. Our resulting solution is an efficient and secure means by which a firewall may be distributed, and obfuscated while maintaining the ability for multiple servers to obliviously evaluate its functionality.  
\end{abstract}

\section{Introduction}
Firewalls are a very common security tool used to protect local networks from threats present in the larger Internet. This is achieved by various strategies, the most basic of which is a packet filter \cite{peltier2006complete}. This type of firewall checks each incoming packet against some set of rules often represented by a \q{blacklist} or \q{whitelist}. With the increasing sophistication of attacks on firewalls from both internal and external threat, there has been an interest to distribute the functionality and management of the firewall \cite{pena2014development,kaur2015programmable,satasiya2016enhanced,filipek2016securing}. While this does allow a large number of machines in the local network to potentially maintain protection without dependence on the functionality of a central server, it also presents the opportunity for individual users inside the firewall to disable or circumvent the rules of their now localized firewall. Furthermore, if the rules of these firewalls exist in plaintext form, as it is often the case in practice, the rules themselves may be considered sensitive information due to a desire to hide the identities and locations of whitelisted external users, or the identities and locations of those blacklisted sources the administration consider as threats.  

If we consider a brief vignette, we may encounter Carl, a corporate employee, who has received some incentive to act in the professional interests of others besides exclusively his employer. Thus, depending on his actions whether they may be passive or active attacks, he can readily and more easily gain access to the corporate system infrastructure. 

\begin{itemize}
	\item In the passive case, he may copy the data contained in the firewall rules. If the firewall rules represent a blacklist, he may learn some information to share concerning the identity or location of entities considered as threats to the corporation. Alternatively, if the firewall represents a whitelist, he may learn and leak information regarding allies or affiliates identities or locations outside the corporate intranet. 
	\item If instead his attacks are more active, he may be able to insert or delete entries from either black or white lists either of which could prove disastrous for the network, its assets, and desired functionalities. 
\end{itemize}

To prevent these attacks, we may argue that firewall rules can be encrypted and access control policies can be used to restrict entities to access firewall configuration files. However, these techniques are not sufficient for the following reasons: 

\begin{itemize}
	\item \emph{Encrypted firewall policies and rules}: When the firewall policies and rules are encrypted with a symmetric key encryption scheme such as AES \cite{AES}, only the person who has the secret key can decrypt these files to access the actual policies and rules. However, during firewall evaluation process for a new network package, the rules have to be decrypted first. This opens the door for attackers to access these policies and rules without knowing the secret key. In addition, when a security breach occurs, the secret key may get exposed, and consequently, so do the encrypted firewall files.  
	
	\item \emph{More restrictive access control policies}: Suppose that mandatory access controls restrict the access to firewall files to the root or a small group of privileged users. Nevertheless, security breaches, exploring unknown vulnerabilities of the network, can bypass access control mechanisms. In addition, human errors and mis-configurations of firewall policies can inadvertently expose sensitive firewall files to other internal users and outside attackers. 
\end{itemize} 
\cbstart
Insider threats to enterprise network security has been a topic of greatly increasing interest among researchers as they seek to address issues of identifying malicious insiders, understanding the limits of the threats they may pose, and the extent of damages which are possible from various thresholds of actions they may take \cite{vetter1997experimental, bowen2009designing}. These researches have progressed into attempts to construct predictive models for identifying malicious inside entities and their attacks \cite{kandias2010insider, schultz2002framework}. The practical utility of such inquiries has been made clear due to the dramatic recent increases of specifically these types of malicious insiders exploiting the access and network privilege with which they have been entrusted. Network security news and information security sites are rife with articles documenting cases of internal breaches and concern over future insider attacks of a similar nature \cite{ruppert2009protecting, dtex2018insider, dunn2018congrats, lindsey2018insider, bandos2018insiders}. 
\cbend

Based on the aforementioned observations, to maximize firewall protection from both internal and external attacks, we need a novel way to manage firewalls. Ideally, firewall rules should be encrypted  and remain encrypted during firewall evaluation process. In addition, we should not rely on one server to manage the firewalls because when a security breach occurs, even encrypted data can get disclosed and firewalls services can be disabled. Therefore, to hide the firewall rules and policies and achieve certain degree of fault-tolerance, we need to utilize a distributed architecture and innovative ways to ``encrypt'' or ``hide'' the firewall rules and policies. Simultaneously, the architecture and firewall hiding method should allow secure and oblivious firewall policy evaluations without decrypting the hidden firewall information. 
\subsection{Our Contributions} 
\begin{figure}
	\centering
	\includegraphics[width=.75\columnwidth]{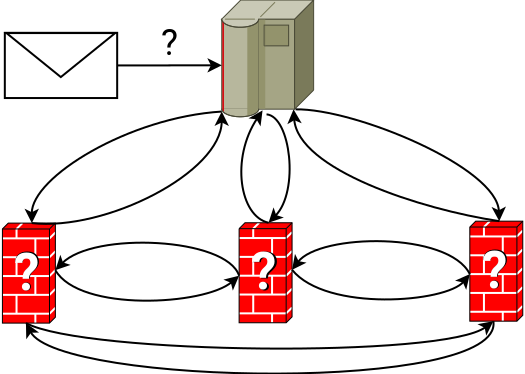}
	\caption{Oblivious firewalls\label{fig:oblivious-firewall}}
\end{figure}
We propose a means by which with relatively low additional overhead, a system is established to 
allow for the rules and functionality of the firewall system to be distributed in such a way that they are information theoretically secure against local inspection, and resistant to tampering due to the techniques we employ. Figure \ref{fig:oblivious-firewall} shows a high level overview of the proposed distributed architecture which requires at least three servers. The firewall is secretly distributed among the servers with either additive or Shamir secret sharing schemes \cite{shamir1979share}; thus, each server alone will not be able to discover any information regarding the firewall. Unlike the previous distributed approaches, our proposed architecture and secure evaluation methods to distributing a firewall does not allow any individual machine to circumvent the firewall.  

Additionally, it could be made resilient to failures as cited as a motivating factor for many other distributed approaches. To achieve this we may allow the servers to continue functioning so long as, out of $m$ total servers, at least some threshold $t$ remain operational. This is possible when implemented, as we suggest, under the Shamir secret sharing scheme. This property of resilience in the face of component failure is a key feature of this scheme. We make use of some previous research, such as this scheme's resilience properties, as well as some novel strategies to achieve these results as will be discussed in Section 
\ref{sec:protocols}.

Furthermore, we employ secure multi-party computation (SMC) techniques to securely and obliviously evaluate the firewall criteria checking function efficiently through the use of a distributed and secret shared Bloom filters. Following the sharing of the Bloom filter constructed from the information a given desired firewall functionality, our proposed methods will generally proceed as follows:
\begin{enumerate}
	\item  During a brief and trusted initialization phase, the filter to represent the firewall is constructed and shared. \item Following this initialization, any gateway receiving a packet invokes a set membership query with respect to the related information. 
	\item The distributed firewall servers, acting as shareholders in the secret sharing scheme, cooperate to calculate shares of the result which is sent to the gateway. 
	\item The gateway reconstructs the result of the distributed multi-party computation and either rejects or forwards the packet. 
\end{enumerate}
The above steps can easily be adopted by many systems with little software modification in their existing routing and firewall protocols. In summary, the desirable properties of our system allows for a local network of arbitrary topology with respect to connection to the outside Internet, to securely, and obliviously, evaluate firewall rules in a distributed manner while maintaining the ability to tolerate some margin of component failure or malicious behavior. We provide analysis concerning the resilience of our scheme against the previously mentioned attacks to demonstrate its merit from the perspective of security, and we analyze the complexity of our approaches to demonstrate its efficiency. Finally, we present some options for various modifications depending on desires for security and the effects these alterations have on efficiency as an interesting trade-off.  Next we formally define the threat models and adversary behaviors that can be prevented by using the proposed techniques. 
The types of adversaries and their attacks that we consider fall into two main categories, passive and active. These types of adversaries have different abilities which present challenges to protocols secure under each. The passive adversary generally follows the prescribed steps of a protocol and maintains a record of the protocol execution for analysis in order to extract additional information from the execution image of the protocol. If any additional information, beyond what can be inferred from the private input and output, the protocol is not secure against this type of adversary. In general, all passive attacks are related to this type of behavior, and there is no need to distinguish between adversaries inside or outside of the local intranet. While the threshold of difficulty for this kind of attack is likely lower for an internal agent, the definition of a passive attack and necessary security measures against it remains the same. 

The active attacks we consider, perpetrated by a malicious adversary, goes further to alter the result of the computations via a wide array of possible actions. In our setting, we consider both these two types of attacks against our attempts to secure the firewall itself, those which are purely passive, and those which are active attacks. Specifically, we consider gaining information from accessing servers which are a part of our proposed system to extract data, monitoring processes and traffic local to a particular machine, or monitoring network traffic attempting to gain information concerning the firewall, disabling servers in our proposed system, modifying protected data on such servers, and reading protected data. Divided into their two categories, the attacks we will consider and make efforts to secure against are the following:
\begin{itemize}
	\item Passive
	\begin{itemize}
		\item Extracting information from local data
		\item Observe traffic passing through a particular machine
		\item Observe network traffic 
	\end{itemize}
	\item Active
	\begin{itemize}
		\item Disable a server participating in the scheme
		\item Read protected data
		\item Modify protected data
	\end{itemize}
\end{itemize}
Under the proposed system architecture, we develop two classes of secure protocols to obliviously manage firewalls. In particular, the first class offers security guarantees against passive adversaries, and the other class can detect malicious activities and the parties involved in these activities, and recover from their activities up to a threshold. Then the malicious activities can be  corrected, and malicious parties can be removed from further participation immediately.

\subsection{Organization} 
The rest of the paper is organized as follows: Section \ref{sec:related} presents foundational information regarding firewalls and more details concerning recent concerns and developments, and
Section  \ref{sec:prelim} discusses the underlying tools needed build up our scheme, including Bloom Filters and Secret Sharing. Then, we present our protocol in detail in Section \ref{sec:protocols}, describing their functionality and procedure along with analysis concerning complexity, security, and correctness. Finally, we conclude by summarizing our contribution in Section \ref{sec:conclusion}, and indicate areas for future work.

\section{Related Work}\label{sec:related}
Firewalls in general have been in existence nearly as long as computer networking. The earliest functions simply involved a router or gateway server maintaining a list of addresses to block or permit. These are so-called \emph{blacklists} and \emph{whitelists} respectively \cite{peltier2006complete}. As sophistication increased divergent interests produced multiple approaches to securing local subnets against threats from the Internet at large. These include rules to block entire domains or sub-domains, as well as a wide array of means by which these rules may be optimized for both security and efficiency. \cite{gouda2004firewall}. While very quick, this method potentially allows for a single point of failure in both terms of network functionality, as well as security. It is possible in this case that an adversary, particularly an internal one, may be able to access the list of permitted or prohibited addresses and use this information outside the desired context or for nefarious purposes. Additionally, if access to the file containing the lists is obtained, it may be possible for an adversary to grant access to other external undesirable, and previously blocked parties, or hinder the ability of a legitimate party to connect. These are important and perhaps critical concerns. 

One means by which a firewall may be represented in terms of a whitelist or blacklist is through the structure of a Bloom Filter \cite{geravand2013bloom}. This type of data structure quickly and efficiently gives set membership in probabilistic terms. The efficiency of this scheme is particularly strong with respect to space. This important topic, on which our work is dependent, will be addressed more thoroughly in Section \ref{sec:prelim}. This data structure has been used in firewall constructions previously to good effect \cite{tarkoma2012theory}.

Additionally, a considerable amount of research is in reliably distributing the functionality of a firewall throughout a network to varying degrees with accordingly varying rates of efficiency and security \cite{broder2004network}. This is motivated by a variety of concerns and for an array of settings from cellular networks to cryptocurrencies and enterprise systems \cite{geravand2013bloom,einziger2017tinyset}. While advantageous from an resilience perspective, or in a situation where restricted network topology is not possible or practical this approach can have risks of its own \cite{ioannidis2000implementing}. If the filter for a particular device or user is local to the system in question, many users may take steps to circumvent this policy. This is much more readily achievable for users with direct access to the system implementing the firewall than if it were a separate dedicated and more secure system in the network infrastructure for the subnet. Our interest in distributing the firewall is to distribute it in an oblivious means, distinct from these approaches, such that the individual machines cannot simultaneously alter the result of the firewall and maintain anonymity. This is discussed in greater depth in the presentation of our proposed protocols in Section \ref{sec:protocols}.

A combined approach allows for the distribution of the firewall, represented by a set of bloom filters \cite{maccari2007mesh}. This is useful especially when network topology is not constrained to a single gateway making the bridge between the local network and the Internet, a common occurrence in mesh networks. In this setting the filters are individually based on the acceptance of packets at each node \cite{maccari2007mesh,neira2008stateful}, and forwarding a packet from one neighbor to another is dependent on the filters associated with the forwarding table. A packet is forwarded if the destination node has whitelisted the source node.  This  approach still allows for potential manipulation of the filters without the ability to detect malicious activities, and is less efficient than more centralized approaches.

Software defined networking has recently shown great promise to ameliorate the issues related to much of networking's inherent hardware dependence. It allows for both hardware abstraction, leading to greater flexibility, as well as a separation between data handling and forwarding and traffic controls \cite{kreutz2015software}. This separation of concerns is certainly generally advantageous from the perspective of flexibility and efficiency, witness the more than a decade ongoing effort to update the infrastructure from IPv4 to IPv6 so the connection is both feasible and reliable \cite{internet2017ipv6}. Even today, adoption is estimated at around only 23\% \cite{goog2017ipv6}. While this has many advantageous properties, it does not inherently impart any greater security to the operations which is a primary motivator for our work. Firewalls implemented under this paradigm may still leak undesirable information, and they may still be maliciously altered in some circumstances. 
\section{Preliminaries}\label{sec:prelim}
For our proposed methods to follow, we require two additional pieces of research upon which our solution is built. These are Bloom Filters, as previously mentioned, as well as secret sharing. These two mathematical structures form the foundation of what we propose. Additionally here we will introduce our discussion of adversary models, some properties of the schemes we employ with respect to these potential adversaries, as well as some introduction to the possibilities for various network topologies.

\subsection{Bloom Filters}\label{sec:bloom}
This probabilistic data structure is eminently beneficial in a setting for which tests of membership in a large set are both frequent, and can tolerate a small to negligible chance of false positives while guaranteeing no false negatives \cite{bloom1970space}. The mathematical underpinning of this data structure relies on hashing, and the probability of hash collisions as well as the probability of the entire vector of values eventually being changed to represent set membership. Constructing a Bloom filter involves generating a vector of bits of length $\beta$. The $\eta$ values to be represented in the filter are each hashed with $\kappa$ distinct hash functions. The result of the hash, modulo $\beta$, is used as an index to the bit array, and the resulting location is set. If the resulting index location was already 1, it is left as 1. Testing for set membership involves calculating the hashes for the object of the query, accessing each location, and verifying that every resulting location is 1. If any location is not 1, the queried object does not belong to the set \cite{tarkoma2012theory}. 

\begin{figure}\centering
	\includegraphics[width=5cm]{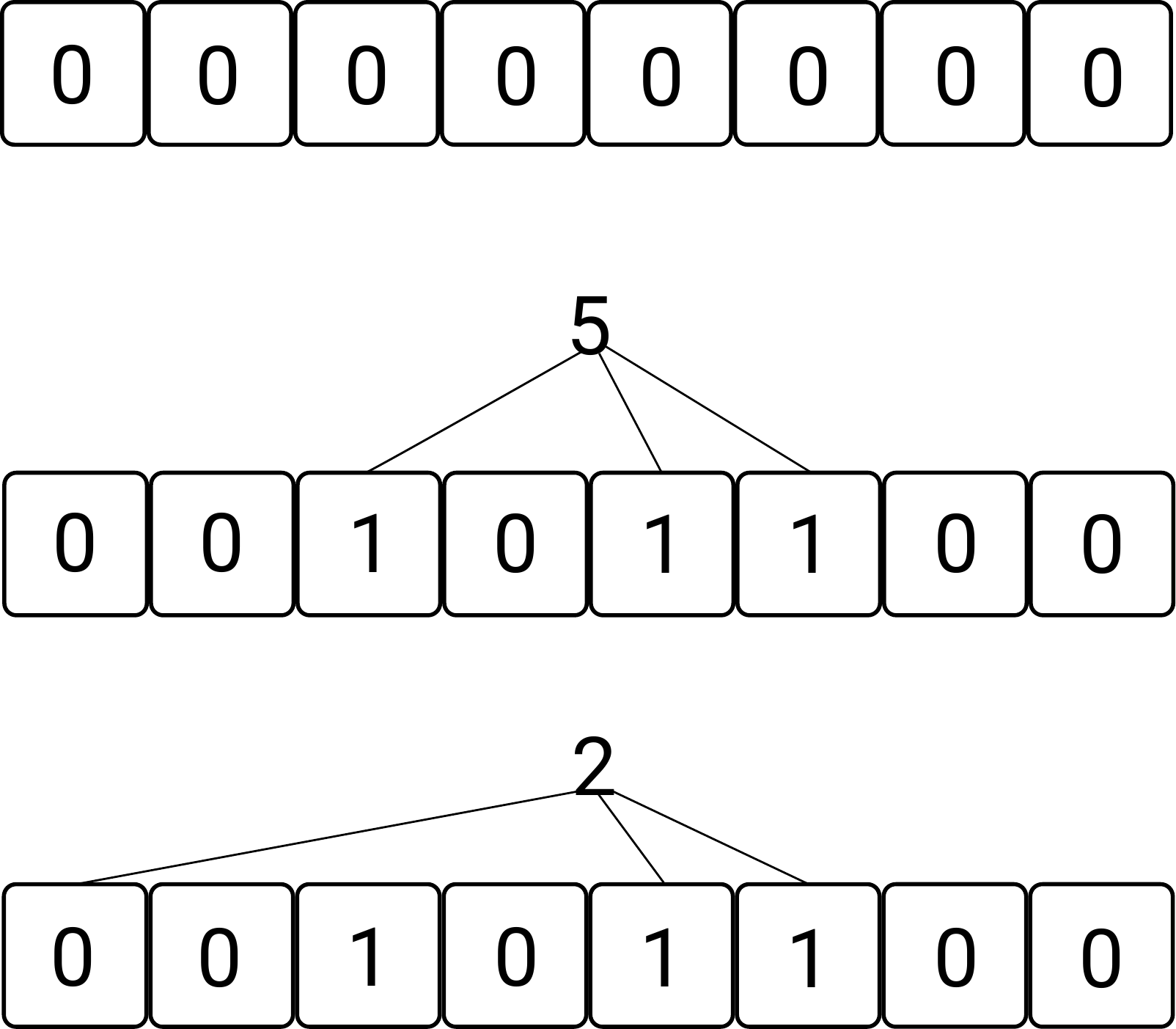}
	\caption{Process of Bloom filter creation, item insertion, and query\label{fig:bloom1}}
\end{figure}

A small and simple example is given in Figure \ref{fig:bloom1}. Here the first vector denotes the initialization of a filter consisting of 8 locations indexed from left to right by values in $\{0\dots7\}$. The second represents the insertion of the element 5. We have hashed the value 5 with three hash functions in this example, FNV132, MD5, and SHA1. The result of each hash function, modulo 8, is 2, 4, and 5, respectively. As can be seen in the figure, each of these locations are set to 1. The filter contains only one element, 5. If we test for set membership with a value 2, we calculate the hash values with the same hash functions, modulo 8, which are 5, 0, and 4, for the same functions, respectively. In the case of indices 5 and 4, the filter contains 1, but for location 0 a 0 is present. All tested locations are not equal to one, therefore the filter correctly indicates that 2 is not a set member. 

The probabilistic nature of this structure is related to the collision probability and the fact that, given a sufficiently large number of elements in the set of interest, and a sufficiently small filter, the rate of collision increases. This results in, eventually, an item which is not in the set testing positive for set membership under a query through a sufficient number of collisions with other elements. Given enough elements, the entire filter may fill with set bits, thus permitting every query to be evaluated as a set member. With appropriately chosen parameters the event of a false positive test for set membership can be made negligible. The relationship between these variables can be derived fairly intuitively. The probability that any individual bit location is still 0 after inserting $\eta$ elements, hashed by $\kappa$ hash functions, in a filter with $\beta$ locations is 
\begin{equation} \label{eq:error-rate}
p=\left(1-\frac{1}{\beta}\right)^{\kappa\eta}\approx e^{-\kappa\eta/\beta}
\end{equation}

Thus the probability of a false positive is approximately $(1-p)^\kappa\approx .6185^{\beta/\eta}$. As discussed in \cite{broder2004network}, the optimal error rate occurs when $p=\frac{1}{2}$ thus a set of parameters for which the following approximate equality holds, constrained by integer values for each parameter, is a set of nearly optimal choices:
\begin{equation}\label{eq:size}
\kappa\approx\frac{\beta}{\eta}\ln 2
\end{equation}


\subsection{Secret Sharing}\label{sec:secret-sharing}
Secret Sharing schemes present the ability to break data up into multiple shares, distribute the shares, and perform computations based on these shares rather than the data, for which a strong concern of privacy exists. This can be used to construct one solution to the classic \q{Millionaire's problem}. 

\subsubsection{Shamir's Secret Sharing Scheme}\label{sec:shamir}
The scheme we focus on for this work is Shamir's secret sharing scheme \cite{shamir1979how}. In this setting, the data is set as the constant term of a polynomial of degree $t-1$ for some threshold $t\geq2$. Arbitrarily many parties $m$ may be involved in the scheme, but to keep complexities minimal, if there is no desire for multiplications, $m$ may be set equal to $t$. If multiplications are desired $m$ may be set to $m=2t-1$. If resilience in either case is desired, in this scheme, it is possible through setting $m>t$ or $m>2t-1$ dependent on the desirability for multiplication or not \cite{bogdanov2007foundations}. 

\paragraph{Constructing shares}
Formally, a secret value, s, is placed within a $t-1$ degree polynomial $f$ of the form:
$$ f(x)=(c_{t-1}x^{t-1}+c_{t-2}x^{t-2}+\dots+c_{2}x^{2}+c_{1}x+s)\mod N $$
in which all constants $c_{t-1}$ through $c_{1}$ are chosen from a uniform random distribution on the domain of the secret, $\mathbb{Z}_N$. For all parties indexed by, $i$, in the range $[1,m]$ the share, for each party is constructed:
$$ [s]_N^{P_i} = f(i)$$
\paragraph{Rebuilding Secrets}
Lagrange Polynomial Interpolation is the method of choice for these operations. Due to our interest being the y-intercept the general formula, 
$$ f(x)=\sum_{j=1}^{m}y_j \prod_{k=1:k\neq j}^{m} \frac{x-x_k}{x_j-x_k}$$
may be somewhat simplified because every variable $x$ in the preceding equation, for our case, is 0.
$$ s= f(0)=\sum_{j=1}^{m}[s]_N^{P_j} \prod_{k=1:k\neq j}^{m} \frac{-x_k}{x_j-x_k}$$
\paragraph{Addition}
In the Shamir Secret Sharing Scheme, addition is \q{free} in that no communication is required between parties, and no lengthy local computations are necessary. For two values, $a$ and $b$, which are shared among the parties, the shared sum is computed by each party $P_i$ locally computing their new share by adding their shares of the secrets $a$ and $b$:
$$ [a+b]_N^{P_i} = [a]_N^{P_i} + [b]_N^{P_i}$$

Addition of a public constant $c$ is affected simply through the addition of this constant into each party's shares
\[[a+c]_N^{P_i}=[a]_N^{P_i}+c\]

\paragraph{Multiplication}

Multiplication of a public constant $c$ is affected simply through the multiplication of this constant into each party's shares
\[[ac]_N^{P_i}=c[a]_N^{P_i}\]

Due to the fact that this scheme is inherently a linear scheme, multiplication of two shared values is a more complex process. If we were to simply multiply the shares as we added them in the previous case, we would get a set of shares representing a polynomial of an order twice that of the what it was previously. This would be disastrous for the system as a whole. For some data, nearly twice the desired number of shares would be required for the reconstruction of the secret. The system would be made up of inconsistent share representations based on polynomials of differing degree. 

In order to achieve multiplication for two shared values $a$ and $b$, and get the product, shared, $ab$, we must compute the product of all the shares. However, it is necessary to also handle some additional communications and calculations by which the final polynomial maintains the same degree as was previously desired, as well as uniform randomness. It is important to note that if multiplication is a desired operation, the number of parties holding shares, $m$, must be greater than or equal to $2t-1$. This is required in order to be able to reduce the order of the polynomial. The degree reduction following the multiplication of the shares is achieved by the sharing of additional randomized information among all those holding shares and then combining the results. This too must be done in a multi-party and secure manner lest intermediate calculations be leaked and provide an adversary an opportunity to learn about the secrets. The introduction and proof of this method is given by Gennaro et al \cite{gennaro1998simplified}.

Each party initially computes the product of their shares $[q]_N^{P_i}\gets [a]_N^{P_i}[b]_N^{P_i}$ which is an intermediate value for each party in this protocol. They each independently, and uniformly randomly construct polynomials of the same degree desired for the system and set the constant, or y-intercept to their intermediate calculated value, $[q]_N^{P_i}$. Each of the local equations will be of the form:
$$ h_i(x)=(c_{t-1}x^{t-1}+c_{t-2}x^{t-2}+\dots+c_{2}x^{2}+c_{1}x+[q]_N^{P_i})\mod N $$
Each party then generates, exactly as the dealer would with a new value to be shared, a set of shares from this polynomial for each of the parties in the scheme. This is done by randomly and uniformly creating shares of the product of the data points by each party and for each party. These must be shared so that a new and similarly uniform random polynomial of the desired degree can be calculated by each peer.
$$\forall i,j \in [1,m], [R]_N^{P{i,j}} =  h_i(j)$$
Players then exchange these shares $[R]_N^{P{i,j}}$ of their intermediately calculated values, sending them from $P_i$ to $P_j$ and keeping the value applicable for its own index, where $i=j$. 

Thus each party gains a set of values from each other party in the scheme. These values must be combined with weights that reflect the reduction of order desired. This recombination vector is applied via a dot product with the values from the other parties and the result is a valid share, of the proper degree, for the product of interest. More details are available in the referenced works, but are beyond the scope of this paper.   

\subsubsection{Additive Secret Sharing}\label{sec:additive}
If the concern for the network is only passive adversaries, a large boost in efficiency may be realized via implementing our proposed solutions to follow under the additive secret sharing scheme. 

This scheme makes use of a different set of mathematical principles to achieve secure multi-party computation, though modular arithmetic still lies at the core of its security. The underlying security is dependent on the fact that adding any value to a uniformly randomly selected value, modulus a value delimiting the group, $N$, is still uniformly random \cite{bogdanov2007securely}. It is therefore impossible to say what the non-random component of the sum was, when considering only the resulting sum. In this context the sum is therefore unconditionally secure since any adversary, unbounded by limits on computational power, can do no better than also simply randomly guess at what the two original values may have been without any means of verification. This security is similar in principle to the unconditional security guarantees of the One Time Pad protocol. The weakness there being in inability to reuse the pad due to leaks in information. In the setting of secret sharing, however, no such concerns exist. For each and every share new uniform random values are selected, used, and distributed, which removes the risk for information leakage from padding value reuse. 

In the context of this scheme some of the operations are more computationally efficient than was the case in the Shamir Secret Sharing Scheme, but, as always, there is a trade-off. With this secret sharing scheme, there is no ability to have less than all the shares that were originally generated involved in the reconstruction of the secrets. While it is fairly trivial  to distribute the shares and reconstruct the secret, all shares are required. This is a special case of the idea previously discussed in Section \ref{sec:shamir}, of a threshold scheme in which $t=m$. Operations in this approach are as follows \cite{bogdanov2008sharemind,bogdanov2013sharemind}.

\paragraph{Constructing Shares}
To create any desired number of shares, $m$, in this scheme, the secret, $s$, to be shared is placed in an equation as follows in which all $[s]_N^{P_1}$ through $[s]_N^{P_{m-1}}$ are uniformly randomly selected integers and $[s]_N^{P_m}$ satisfies the equation, all in $\mathbb{Z}_N$:
$$[s]_N^{P_m}=s-[s]_N^{P_1}-[s]_N^{P_2}-\dots -[s]_N^{P_{m-1}} \mod N$$

\paragraph{Rebuilding Secrets} Rebuilding secrets in this case is also trivial since it can clearly be seen from the preceding all one need do is sum all the $m$, shares held by the parties modulus the same integer $N$:
$$s=\sum_{i=1}^m[s]_N^{P_i} \mod N$$
This is the greatest area of difference for our efficiency concerns between the two schemes. Additive secret sharing requires only additions modulo $N$ to rebuild secrets while Shamir's scheme requires polynomial interpolations. Interpolations of course requiring, more operations, growing at a higher asymptotic complexity, and requiring more complex operations. Once again, however, we emphasize the loss of resilience as well as being secure against only passive attacks.  

\paragraph{Addition}
Since this scheme simply makes use of addition modulo $N$, it is possible for each party in the scheme to acquire the share for the sum of two secrets by simply adding, modulo $N$, their local shares of the two secrets. This can be seen for the two secrets $a$, and $b$, here via some arithmetic manipulation:
\begin{align*}
a+b&=([a]_N^{P_1}+[a]_N^{P_2}+\dots +[a]_N^{P_m})\\ &+([b]_N^{P_1}+[b]_N^{P_2}+\dots +[b]_N^{P_m})\mod N
\end{align*}

$$a+b=([a]_N^{P_1}+[b]_N^{P_1})+\dots +([a]_N^{P_m}+[b]_N^{P_m})\mod N$$
\paragraph{Multiplication}
This is affected among three parties through the generation, distribution, and use of multiple additional uniform random values from the group. They are combined and shared among the parties in a multi-step process which creates a complex, but secure, uniform random sharing of the desired product among the parties. The protocol given in the work of Bogdanov \cite{bogdanov2007securely,du2001protocols}, is as follows in Algorithm \ref{alg:addmul} for two secrets $u$, and $v$, with shares $[u]_N^{P_1},[u]_N^{P_2},[u]_N^{P_3}$ and a similar set for $v$. 

\begin{algorithm} 
	\caption{\newline SMM$(\langle P_1, [u]_N^{P_1}, [v]_N^{P_1}\rangle, \langle P_2, [u]_N^{P_2}, [v]_N^{P_2} \rangle, \langle P_3,[u]_N^{P_3}, [v]_N^{P_3} \rangle)\newline\rightarrow  (\langle P_1, [uv]_N^{P_1}\rangle, \langle P_2, [uv]_N^{P_2} \rangle,\langle P_3,[uv]_N^{P_3} \rangle) $\label{alg:addmul}}
	Round 1\\
	\hspace{.5cm}Party 1 generates $r_{12},r_{13},s_{12},s_{13},t_{12} \in_R \mathbb{Z}_N$\\
	\hspace{.5cm}Party 2 generates $r_{21},r_{23},s_{21},s_{23},t_{23} \in_R \mathbb{Z}_N$\\
	\hspace{.5cm}Party 3 generates $r_{31},r_{32},s_{31},s_{32},t_{31} \in_R \mathbb{Z}_N$\\
	\hspace{.5cm}All values $*_{ij}$ are sent from $P_i$ to $P_j$\\
	Round 2\\
	\hspace{.5cm}Party 1 Computes:\\
	\hspace{1cm}$\hat{a}_{12}=[u]_N^{P_1}+r_{31}$ \hfill $\hat{b}_{12}=[v]_N^{P_1}+s_{31}$\\
	\hspace{1cm}$\hat{a}_{13}=[u]_N^{P_1}+r_{21}$\hfill $\hat{b}_{13}=[v]_N^{P_1}+s_{21}$\\
	\hspace{.5cm}Party 2 Computes:\\
	\hspace{1cm}$\hat{a}_{23}=[u]_N^{P_2}+r_{23}$ \hfill $\hat{b}_{23}=[v]_N^{P_2}+s_{12}$\\ \hspace{1cm}$\hat{a}_{21}=[u]_N^{P_2}+r_{32}$\hfill$\hat{b}_{21}=[v]_N^{P_2}+s_{32}$\\
	\hspace{.5cm}Party 3 Computes:\\
	\hspace{1cm}$\hat{a}_{31}=[u]_N^{P_3}+r_{23}$\hfill$\hat{b}_{31}=[v]_N^{P_3}+s_{23}$\\ 
	\hspace{1cm}$\hat{a}_{32}=[u]_N^{P_3}+r_{13}$\hfill$\hat{b}_{32}=[v]_N^{P_3}+s_{13}$\\
	\hspace{.5cm}All values $*_{ij}$ are sent from $P_i$ to $P_j$\\
	Round 3\\
	\hspace{.5cm}Party 1 computes:\\
	\hspace{1cm}\vspace*{-1.5\baselineskip}\begin{align*}
	c_1&=[u]_N^{P_1}\hat{b}_{21}+[u]_N^{P_1}\hat{b}_{31}+[v]_N^{P_1}\hat{a}_{21}+[v]_N^{P_1}\hat{a}_{31}\\ &-\hat{a}_{12}\hat{b}_{21}-\hat{b}_{12}\hat{a}_{21}+r_{12}s_{13}\\ &+s_{12}r_{13}-t_{12}+t_{31}
	\end{align*}\\
	\hspace{1cm}$[uv]_N^{P_1}=c_1+[u]_N^{P_1}[v]_N^{P_1}$\\
	\hspace{.5cm}Party 2 computes:\\
	\hspace{1cm}\vspace*{-1.5\baselineskip}\begin{align*}
	c_2&=[u]_N^{P_2}\hat{b}_{32}+[u]_N^{P_2}\hat{b}_{12}+[v]_N^{P_2}\hat{a}_{32}+[v]_N^{P_2}\hat{a}_{12}\\ &-\hat{a}_{23}\hat{b}_{32}-\hat{b}_{23}\hat{a}_{32}+r_{23}s_{21}\\ &+s_{23}r_{21}-t_{23}+t_{12}
	\end{align*}\\
	\hspace{1cm}$[uv]_N^{P_2}=c_2+[u]_N^{P_2}[v]_N^{P_2}$\\
	\hspace{.5cm}Party 3 computes:\\
	\hspace{1cm}\vspace*{-1.5\baselineskip}\begin{align*}
	c_3&=[u]_N^{P_3}\hat{b}_{32}+[u]_N^{P_3}\hat{b}_{23}+[v]_N^{P_3}\hat{a}_{13}+[v]_N^{P_3}\hat{a}_{23}\\ &-\hat{a}_{31}\hat{b}_{13}-\hat{b}_{31}\hat{a}_{13}+r_{31}s_{32}\\ &+s_{31}r_{32}-t_{31}+t_{23}
	\end{align*}\\
	\hspace{1cm}$[uv]_N^{P_3}=c_3+[u]_N^{P_3}[v]_N^{P_3}$\\
	
\end{algorithm}
The proof of correctness is verbose, but straightforward, we would direct the reader to the thorough treatment it is given in \cite{bogdanov2007securely}. The communication complexity of this protocol is \compBogThree\ bits in three rounds among the three parties according to the referenced paper and analysis. However, in consonance with our definition concerning a computational round, this could be considered as requiring two rounds.

\subsubsection{Unbounded Fan-in Multiplication}\label{sec:fanmult}
It is possible to compute products an arbitrary number of secret shared group elements in constant rounds, for either of the two preceding schemes, through the methods introduced my Bar Ilan and Beaver \cite{bar1989non}. For a set of $k$ elements which should all be multiplied together, $k+1$ random shared elements must be generated, and have their shared multiplicative inverse calculated. Thus $[r_0]_N^{P_j}\dots[r_k]_N^{P_j}$ as well as $[r_0^{-1}]_N^{P_j}\dots[r_k^{-1}]_N^{P_j}$ should be shared among all the participating parties $P_j$. For the set of elements for which the product is desired $[s_1]_N^{P_j}\dots[s_k]_N^{P_j}$ the parties need to cooperate in calculating and reconstructing 
\[ s'_i=[r_{i-1}]_N^{P_j}[s_i]_N^{P_j}[r_i]_N^{P_j} \]
for all $1\leq i\leq k$. Locally the parties can compute the product of all these $s'_i$ such that
\[s'=\prod_{i=1}^{k}s'_i\]
Finally one additional cooperative calculation yields shares of the desired product
\[ s=[r_0^{-1}]_N^{P_j}s'[r_k]_N^{P_j} \]
In total such a product requires generating 2(k+1) random values, evaluating 3k+2 binary input products, and cooperating in 2k+1 reveal operations in 5 rounds. To go into more detail than counts of secure multi-party protocol invocations a selection of a scheme is required. Differences in complexity for all aspects, local computations, communications, and round complexity exist between Shamir's secret sharing scheme and additive secret sharing.

\section{The Proposed Protocols} \label{sec:protocols}
In this section, we propose our solutions by which a firewall can be secured in an information-theoretic sense and distributed across multiple servers that evaluate the firewall functions quickly and efficiently while maintaining a higher level of security than previously proposed systems. The proposed solutions  can also be expanded to include resilience in the face of some subset of servers failing and being  additionally secure against attempts to manipulate the firewall via the primitives and nature of the underlying mathematical constructs. In the descriptions to follow, we have assumed a blacklist approach to the firewall construction due to the possible but unlikely event of a false positive test result being more tolerable in general for a firewall.

The proposed protocols can be classified into three categories according to different aspects of firewall
management: (1) firewall initialization, (2) firewall rule evaluation and (3) firewall rule or policy update. 

\begin{itemize}
	\item \emph{Firewall initialization}: Given a false positive error rate, the size of the blacklist, the 
	initialization protocols can decide the Bloom filter size and the number of hash functions. Then 
	the Boom filter representation of the blacklist will be produced. Depending on the adversary model,
	either additive or Shamir secret sharing scheme, as discussed in Section \ref{sec:secret-sharing},
	will be used to generate secretly shared Bloom filter,
	and each share is provided to one of the designated servers for oblivious firewall management and 
	evaluation. Thus, each server has shares of the Bloom filter representing the rules of the firewall, and they are collectively ready for firewall rule evaluations. This initialization of generating secret shares of the Boom filter is assumed to be a trusted process, and  after that, we consider various threats to the security of the proposed methods to obliviously manipulate and evaluate the firewall policies and rules. 
	\item \emph{Firewall rule evaluation}: We developed several protocols for firewall rule evaluation
	with various security guarantees.
	In the proposed protocols, we assume that the firewall has been initialized, that is, 
	the information of interest, such as IP addresses on 
	the blacklist have already been hashed and a Bloom filter has been constructed and 
	exist as shares across the servers to be involved in the firewall evaluation. 
	\item \emph{Firewall rule or policy update}: When firewall rules and policies are updated, 
	the Bloom filter representing this firewall also needs to be updated according. In addition, secret shares of 
	the Bloom filter need to be updated as well at each server. We developed a secure protocol to 
	perform these actions to update our oblivious firewall.
\end{itemize}

For the rest of the paper, our proposed schemes are described with several key system parameters shown in Table \ref{tab:notations}. Before providing the details for implementing a secured and distributed firewall evaluation function among independent servers, we next clarify the network topologies assumed in this work.  

\begin{table}[ht]
	\begin{tabular}{p{.5cm}p{7cm}}
		\hline
		\hline
		$\eta$ & The number of addresses to add to the Bloom filter\\
		$\beta$ & Number of bit locations in the Bloom filter\\
		$\kappa$ & Number of hash functions used in checking set membership via the Bloom filter\\
		$t$ & Threshold from the Shamir secret sharing scheme\\
		$m$ & Number of parties involved in the secret sharing scheme\\
		$N$ & Prime modulus for the secret sharing scheme \\
		\hline
		\hline
		\vspace*{.001cm}
	\end{tabular}
	\caption{Common notations}
	\label{tab:notations}
\end{table}

\begin{figure}[]
	\centering
	\vspace{.5cm}
	\begin{minipage}{.47\columnwidth}
		\centering
		\includegraphics[width=\linewidth]{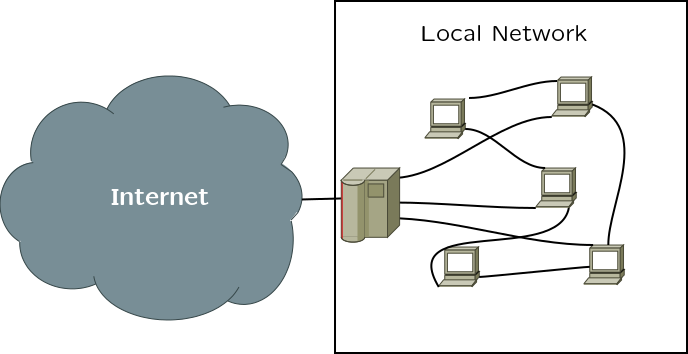}
		\caption{Constrained Network Topology\label{fig:single}}
	\end{minipage}\hspace{.5cm}\begin{minipage}{.47\columnwidth}
		\centering
		\includegraphics[width=\linewidth]{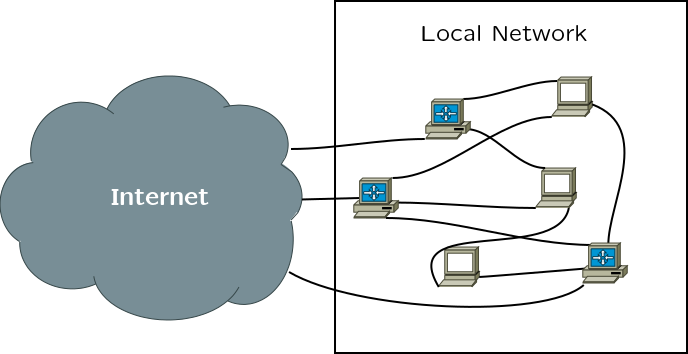}
		\caption{Unconstrained Network Topology\label{fig:many}}
	\end{minipage}\vspace{-.5cm}
\end{figure}

\subsection{Network Topology}
Two general types of situations exist with respect to firewall concerns. Settings in which the network topology is constrained such that there exists a single gateway server to the wider Internet, such as which is depicted in Figure \ref{fig:single}. The other situation is for  less control than a pure centralized topology, but it is centralized up to an arbitrary number of links between the local network and the rest of the Internet. This is depicted in Figure \ref{fig:many}. For our following proposed approaches, unlike either of the previously discussed cases from Section \ref{sec:related}, the structure of the firewall is distributed in a secret shared format. The control function is evaluated via secure multi-party computation techniques, and either topology is acceptable in our scheme.  

\subsubsection{Centralized Topology}
In the centralized or constrained topology setting, the firewall exists on the main server acting as a gateway. With every packet arriving, the sole bridge between the local and external networks conducts the source checking against the rules of the firewall. The result of this check leads to the packet's forwarding or rejection. The rules in this case normally exist within the settings of this single machine and are thus potentially and easily manipulatable, particularly by a malicious insider. Once the malicious behaviors have been carried out, it may be difficult for parties within the network to 
identify the alteration which represents a serious security risk to the local network \cite{peltier2006complete}. 

As mentioned previously, there may be additional motivations to keep the rules of the system itself secret from most parties after the system has been initialized. Disclosure of these rules may result in undesirable leaks of information regarding people and entities outside of the local organization. For our protocol to follow, this gateway would invoke the protocols among the servers who would evaluate the firewall function and all send their shares of the result back to the gateway. The gateway can then reconstruct the result from combinations of the shares and both identify malicious behavior by the servers and permit or reject network traffic in accordance with the firewall rules. 

\subsubsection{Distributed Topology}
In a situation without the benefit of a constrained topology, the firewall must be distributed by some means. If many machines may have access to the external larger Internet, this mesh must have tools to consistently, efficiently and reliably maintain the functionality of the firewall in a distributed manner. This is achieved in many cases by associating the firewall rules with the forwarding tables for other network nodes \cite{broder2004network}. However, these existing methods dramatically increase the risk of information leakage regarding firewall rules, alteration and inconsistency of the rules among each machine in the network. 

For our proposed solution to follow, each machine on the fringe of the network, connected both internally and to the external network, can act similarly to the constrained case previously described. Each of these machines, serving as a gateway, can invoke the the protocols among the servers sharing the filter who all return shares of the result of the firewall evaluation according to our forthcoming protocols. Again, the invoking machine would use combinations of the shares to rebuild the result to identify malicious activity and control packet flow. 

\subsection{Initialization}
The main challenge during the firewall initialization process is to determine the Bloom filter and 
the number of hash functions needed to construct the filter. 
Equations \ref{eq:error-rate} and \ref{eq:size} given in Section \ref{sec:bloom} can help the decision making 
in this regard. Since the size of the blacklist $\eta$ is known and the probability of a false positive is approximated
by $0.6185^{\beta/\eta}$, fixing the false positive rate will lead to the size of the Bloom filter $\beta$. 
Consequently, the number of hash functions $\kappa$ can be calculated by Equation \ref{eq:size}. 
For example, suppose $\eta $= 1,000,000 (one million)and the false positive rate is 0.001. Then $\beta \approx 14.5$
million bits. As a result, the number of hash functions is about $14.5 \times \ln 2 \approx 10$.  
As for the hash functions, we can adopt universal hash in the form of $hash(x) = ax+b \mod q$, where $q$
is the smallest prime bigger than $\beta$, and $a$ and $b$ are randomly selected from $\{1, \ldots, q-1\}$.   

\begin{algorithm}[!htbp]
	\caption{\newline firewallInit$(\langle Admin, addr_i\rangle, \langle P_1,\perp\rangle \dots \langle P_m,\perp\rangle)\newline\rightarrow  (\langle Admin, \perp\rangle, \langle P_1, [\phi_j]_N^{P_1}\rangle\dots \langle P_m, [\phi_j]_N^{P_m} \rangle) $\label{alg:fireinit}}
	\DontPrintSemicolon
	\SetKwInput{KwData}{Input}
	\SetKwInput{KwResult}{Output}
	\KwData{IP addresses ($addr_i$) or other information of interest, for $i=1\dots\eta$}
	\KwResult{$\phi_j$ for $j=1\dots\beta$ is shared among the $m$ parties }
	Admin
	\begin{enumerate}[label=(\alph*)]
		\item $\phi_j=0$, for $j=1\dots\beta$
		\item $\phi_j=1, \forall j\in \{hash_q(addr_i)\mod \beta $ and  $\forall q \in {1\dots\kappa},i\in{1\dots\eta}\}$
		\item $genShares(\phi_j)$, for $j=1\dots\beta$
		\item send $[\phi_j]_N^{P_i}$ to $P_i$, for $i=1\dots m$ and $\ j=1\dots\beta$		
	\end{enumerate} 
\end{algorithm}

During a trusted initialization phase, the addresses forming the blacklist are hashed with the prescribed functions, and the Bloom filter is constructed. This vector of bits is secret shared using the methods introduced in Section \ref{sec:prelim} and sent to each of the shareholder parties to participate in the scheme. Once these two steps have been completed the online operation of the system may commence. The main steps of the 
initialization process are presented in Algorithm \ref{alg:fireinit}.

\subsection{Secure Distributed Firewall Evaluation}
As mentioned before, we developed two secure protocols to evaluate firewall rules obliviously 
distributed among $m$ servers: $P_1$, \ldots, $P_m$. 
In this section, we only discuss the first one. The main evaluation condition behind this protocol is 
based on the following condition determined by the construction of the Bloom filter:
\begin{itemize}
	\item The IP address (\emph{addr}) of the incoming message is on the blacklist if the sum of the values stored in the 
	Bloom filter locations indexed by the $\kappa$ hash functions with \emph{addr} as their input is equal to $\kappa$. 
\end{itemize}
If the above condition is true, the package will be blocked. The design of our first protocol is to ensure 
the condition can be verified without disclosing the actual content of the Bloom filter. 
The key steps of  the protocol are presented in Algorithm \ref{alg:firesum}.

\begin{algorithm}[!htbp]
	\caption{\newline firewallEval$(\langle Gateway, addr\rangle, \langle P_1,\perp\rangle \dots \langle P_m,\perp\rangle)\newline\rightarrow  (\langle Gateway, \sigma\rangle, \langle P_1, \perp\rangle\dots \langle P_m, \perp \rangle) $\label{alg:firesum}}
	\DontPrintSemicolon
	\SetKwInput{KwData}{Input}
	\SetKwInput{KwResult}{Output}
	\KwData{IP address (addr) or other information of interest, parties should be in possession of shares of the firewall represented as a Bloom filter, this is a vector $[\phi_j]_N^{P_i}$, for $j=0\dots \beta-1$ and $i=1\dots m$,
		Public info: $N$ the modulus of the secret sharing scheme}
	\KwResult{$\sigma$ is revealed by the Gateway }
	Gateway
	\begin{enumerate}[label=(\alph*)]
		\item $addr\gets$ IP address (or other info of interest) extracted \\ from packet
		\item Send $addr$ to $P_i$, for $i=1\dots m$		
	\end{enumerate} 
	\nl$P_i$ for $i=1\dots m$ 
	\begin{enumerate}[label=(\alph*)]
		\item $[\sigma]_N^{P_i}=\sum_j[\phi_j]_N^{P_i}: j\in \{hash_q(addr)\mod \beta, \forall q \in {1\dots\kappa}\}$
		\item Send $[\sigma]_N^{P_i}$ to Gateway
	\end{enumerate} 
	\nl Gateway 
	\begin{enumerate}[label=(\alph*)]
		\item $reveal(\sigma) $ for all valid combinations of $t$ out of $m'$ shares within 
		a specific time window, where $m' < m$  
		\item if $m' > t$ and all ${m\choose t}$ reveals do not agree: flag malicious behavior
		\item if the majority of all the revealed $\sigma$ values are equal to $\kappa$: block packet 
		\item else: forward packet
	\end{enumerate}
\end{algorithm}
\subsubsection{Key steps and correctness analysis}
Note that the parameter $N$ denote the size of the secret shares. In this protocol, in order to represent the sum
of the values from locations indexed by the $\kappa$ hash functions, $N$ needs to be bigger than $\kappa$. 
In step 1, the gateway receives a packet from which the IP address or other information of interest  is extracted. 
The IP address is sent to the set of parties (the servers who manage the firewall) 
implementing the distributed secret sharing scheme. In step two, the parties hash the received address with all $k$ hash functions modulo $\beta$, and this set of calculated values are the indices to the locations of the
secretly shared Bloom filter, and the values (i.e., secret shares) stored in these locations should be summed. 
This summation is computed, and the result is sent back to the Gateway.

In step 3, the Gateway reconstructs 
the result of the computation and compares it against $\kappa$. 
Due to the nature of the Bloom filter,
this result will be equal to $\kappa$ if and only if every location in the summation was 1, which is exactly the criteria 
indicating set membership according to the Bloom filter construction. 
Since network faults may occur and malicious adversaries may disable one or more servers among the $m$ servers, 
the gateway could only receive $m'$ shares within an expected time window.
The additional combinations and criteria are directed at security concerns for the proposed protocol which
is addressed in greater detail in Section \ref{sec:sec1}.

\subsubsection{Complexity analysis}
In this setting, the Gateway sends the data of interest to the set of servers forming the shareholding parties in the secret sharing scheme. We assume that this is an IP address in the context of our presentation, and we will continue that common and useful assumption in the context of this analysis. As we note in Section \ref{sec:related}, even though we are more than 10 years out from the acceptance of the IPv6 standards, IPv4 is still strongly in the majority for traffic, thus we assume IPv4 addresses of 32 bits. In the second step of the protocol, all $m$ parties send the result of their summations to the Gateway thus the required communications will be $m\log_2N$. This send and receive cycle constitutes one computational round, and completes the communication requirements for this protocol. The total communication complexity 
is $m(32+\log_2N)$ bits in a single round. 

\subsubsection{Proofs of security for passive attacks}\label{sec:sec1}
We define our notions of security in consonance with the proposed definitions and conventions of Goldreich \cite{goldreich2004applications}. As discussed at length there, a protocol is secure in the computation of a desired functionality given that the view of the execution of the protocol, or execution image, is indistinguishable from that generated by a simulator. Here we define our security in information theoretic terms for passive attacks as described previously. These attacks include attempts to extract information from what data may be visible on a compromised server, observing traffic on a compromised server, and observing visible network traffic. 

In general, with respect to the last point, we assume that the servers are in one of two situations, they are all on different local subnets, or all the servers communications are encrypted in transit. Under these circumstances, the 
attackers can only observe encrypted internal network traffic. Thus, no information regarding the Bloom filter or 
the firewall rules and policies are leaked to the attachers.  With respect to the two other concerns listed, we provide the following analysis and arguments for security.

Given a functionality $f(x,y)$ operating on the inputs $x$ and $y$ and a protocol $\Uppi$ implementing this functionality,
then an execution image for $\Uppi$  is denoted by $\{\textsc{view}_1^{\Uppi}(x,y)\}_{x,y\in\{0,1\}^*}$ for party $P_1$. 
What is needed to prove security of $\Uppi$ is an algorithm $S_1$ which, given the public information as well as the necessary private information from $P_1$, $P_1$ is able to produce a simulated execution  image of $\Uppi$,
denoted by $\{ S_1(x, f_1(x,y) )\}_{x,y\in\{0,1\}^*}$ for some simulator algorithm $S_1$.
If there is an equivalence between the simulated and real execution images in an information theoretic sense, 
then the protocol is secure with this guarantee for passive adversaries. Thus, the goal is to demonstrate the following equivalence which is given for a party $P_1$ though, in general, it would need to be demonstrated for all parties:
\[\{ S_1(x, f_1(x,y) )\}_{x,y\in\{0,1\}^*}\equiv \{\textsc{view}_1^{\Uppi}(x,y)\}_{x,y\in\{0,1\}^*}\]
Since our protocol is symmetric; that is, all parties $P_1, \ldots, P_m$ who are responsible for 
obliviously managing the firewalls perform the same operations, 
it is sufficient to show the equivalence from one party's perspective to prove the protocol is secure for all parties. 

The view of the servers in this situation is comprised of two different components. First they receive shares of the Bloom filter, these shares disclose no information regarding the filter itself, or the firewall rules it represents. This is immediate from the underlying secret sharing scheme. They additionally receive IP addresses which are queried for membership in the set comprising the firewall rules. The only use of this information is in generating the indices for share summation. Summing the shares, or any other local operation performed, yields no information to a shareholder without bounds on available computational power. This too is immediate from the properties of the secret sharing scheme. 
Thus, a trivially implementable simulator for this view consists of sending a random string of 32 bits
representing an IP address to query on the system. The servers are expecting arbitrary 32 bit strings and will proceed in their calculations revealing no information among themselves without malicious behaviors. This is the property we wished to demonstrate for passive adversaries. 

\paragraph{Protocol simplification} If security concerns are only against passive attacks, one may eliminate revealing many combinations of shares at steps 3(a) and 3(b) of Algorithm \ref{alg:firesum}.In this case, we can use additive secret sharing as opposed to Shamir's scheme and immediately realize a considerable performance improvement as alluded to in Section \ref{sec:additive}. With the change of secret sharing scheme, the step 3(b) can be eliminated, and as for the \emph{reveal} function at step 3(a), the Gateway only needs  to execute one summation among all the shares received from the servers and one comparison to make a decision, a very fast and efficient operation. Therefore, the final steps of the protocol can be simplified, but we have chosen to present the more complex and more secure approach earlier as we will presently explain.

\subsubsection{Malicious adversaries}
Malicious adversaries may do anything a passive adversary can do and more. For example, a malicious adversary may deviate from the protocol or manipulate their intermediate calculations. These deviations we wish to at least detect, and possibly identify the malicious party as well. Specifically, the active attacks we consider are (1) disabling a participating server, (2) reading data which is normally expected to be protected, and (3) modifying data which is normally expected to be protected. 

\paragraph{A combinational approach}With respect to the first concern listed, we would draw attention to the fact that this is \q{built in} to the Shamir's secret sharing scheme provided that $m>t$, since any set of $t$ shares can be used to reconstruct the secret. With respect to the other two concerns we provide the following analysis and arguments for security. 

Shamir's secret sharing scheme affords us the opportunity to achieve detection and identification of malicious adversaries relatively efficiently due to the fact that it is a threshold secret sharing scheme. If a sufficiently large number of parties, greater than the threshold, participate in the scheme, it is straightforward to detect malicious manipulation of the shares, or, with more parties, identify a  malicious adversary. As we have discussed in Section \ref{sec:shamir}, Shamir's scheme uses a threshold number of parties to reveal secret shared values, while an arbitrarily large number of parties may participate in the scheme. If the number of parties $m$ is set such that $m\geq t+1$, then a malicious adversary may manipulate their share of the computational result to change the revealed value, only when their share of the result is used in the revealing process. Thus, for a set of 4 parties, a threshold of three, a malicious $P_1$, and a shared value $s$, the following equivalence relations, meaning the output of the reveal function being incorrect, hold for the reveal functionality when $P_1$ has altered $[s]_N^{P_1}$. 


\begin{eqnarray*}
	\text{reveal}\left([s]_N^{P_1},[s]_N^{P_2},[s]_N^{P_3} \right) & \equiv & \text{reveal}\left([s]_N^{P_1},[s]_N^{P_3},[s]_N^{P_4}\right) \\
	& \equiv & \text{reveal}\left([s]_N^{P_1},[s]_N^{P_2},[s]_N^{P_4}\right)
\end{eqnarray*}
However, the other available combination of shares is not equal to the other two, specifically

\[\text{reveal}\left([s]_N^{P_1},[s]_N^{P_2},[s]_N^{P_3}\right)\not \equiv \text{reveal}\left([s]_N^{P_2},[s]_N^{P_3},[s]_N^{P_4}\right)\]
by the simple and direct principles of polynomial interpolation on which this functionality depends. 
Therefore, in this case it is possible to detect malicious behavior due to the fact that all combinations of shares used for reconstruction of the secret do not agree. There is a problem still. The majority of the revealed values are all influenced by the malicious party's actions.

When the difference between $m$ and $t$ grows wider, e.g., $m\geq 2t+1$, revealing shared values becomes more treacherous for a malicious adversary. In this case, the majority of revealed combinations will agree on the correct value, and identifying the malicious party becomes very easy. In the minimal case, where $m=2t+1$ there are $  {2t+1 \choose t}$ combinations possible for revealing the shared value, and only ${2t \choose t-1}$ of them will be able to be influenced by any malicious party. Therefore, the following expression gives the exact fraction of combinations of shares to be used in polynomial reconstruction which it will be possible for a malicious adversary to influence:
\[\frac{{2t\choose t-1}}{{2t+1\choose t}}=\frac{\frac{2t!}{(t-1)!(t+1)!}}{\frac{(2t+1)!}{t!(t+1)!}}=\frac{t}{2t+1}\]
Additionally, in general, a set of $x$ malicious parties, by the same reasoning, the maximum number of shares 
that can be  influenced by the malicious parties is defined by: 
\[ \sum_{i=1}^{\min(x,t)} {x\choose i}{m-x \choose t-i}\]
Thus, in order for the honest adversaries to be able to agree on the correct value by a majority agreement scheme, 
such as that represented in the Byzantine General's problem, a majority of the share combinations must recombine to form the correct solution. Therefore, the following inequality should be preserved to guarantee the correctness of the protocol 
against actively malicious adversaries:
\[ {m \choose t}>2\sum_{i=1}^{\min(x,t)} {x\choose i}{m-x \choose t-i} \]

In a specific case, where $t=3$ and $m=2t+1=7$, there are 35 combinations valid for reconstructing the shared secret. It is evident that a malicious entity can influence 15 combinations for rebuilding the secret, while the majority 20 of the share combinations will remain uninfluenced. Thus not only will a malicious party's activity be detectable, the party is easily identifiable through the minority of reconstructed values consisting of those with exactly one member in common across all combinations of shares used in their reconstruction, specifically the malicious party. 

Suppose $m'$ is the number of messages received by the gateway from the $m$ servers, where $t \le m' \le m$. 
The following conditions specify when malicious behaviors can be detected or prevented:
\begin{itemize}
	\item Malicious behavior detection:  When there are at least $t$ shares from honest servers, the detection of malicious behaviors is possible.
	\item Correctness guarantee (or malicious behavior prevention):
	When ${m' \choose t}>2\sum_{i=1}^{\min(x,t)} {x\choose i}{m'-x \choose t-i}$ and $x < m'$, 
	malicious behaviors, including disabling the servers and modifying the shares, will not influence the 
	correctness of the protocol. 
\end{itemize} 
\cbstart
\paragraph{Berlekamp-Welch Error Correcting Algorithm}
The preceding approach is sufficient for small sets of parties with few to no adversaries and can be reasonably efficient for this setting. If large sets of parties are involved the means which we have proposed for checking for malicious behavior become cumbersome and prohibitive in complexity. When large sets of parties are involved another approach, will be more beneficial. We do not address the details of this alteration in any great depth, but do present the intuition of the additional steps for consideration of this potential case. 

We base the additional steps on the well known, resilient, and efficient Berlekamp-Welch algorithm for error correction \cite{welch1986error}. Since the parties already hold shares of a polynomial, all that is required is a different set of steps at the server holing the packet in question. Rather than rebuilding the shares to reclaim the secret as is normally done, the shares received are used to construct a linear system which is solved locally. This allows for the correct reconstruction of the polynomial underlying the shares representing the filter evaluation, which is equivalent to revealing the shared secret, but with some extra benefit due to special properties of the system. The identification of the error laden points in the interpolation is possible thereby allowing the honest parties to root out the adversary in their midst. This does not hold in all cases for any values of $m$ the number of parties which are shareholders in the scheme, and $t$ the threshold of the scheme, but only for at most $\frac{m-t+1}{2}$ malicious shareholders in the scheme. This is achieved by identifying the shares, really points on a polynomial is Shamir's scheme, which are causing the majority of other points to not lie on the interpolated polynomial. The local complexity of this operation is straightforward in $\mathcal{O}(m^3)$ which is obviously a large improvement over $\mathcal{O} {m\choose t}\approx \mathcal{O}(m!)$ for large groups of parties. Lower complexities than $\mathcal{O}(m^3) $ are possible through other methods.     
\cbend 
\paragraph{Byzantine Agreement}
In the previous presentation of our protocols the ultimate decision to block or forward a packet was made based on the computations of the Gateway alone. While this is a very efficient means, it may not meet the security requirements for all interested parties. We thus further extend the security of our protocol by requiring that the decision to accept or reject the packets be a multi-party computation as well. We achieve this by altering the final step of the protocols, as previously given, to send the shared computation result to not only the Gateway, but every other party involved in the scheme. This necessarily increases the complexity by an additional $m(m-1)\log_2N$ bits, but allows for each of the servers along with the Gateway to rebuild the result of the computation. Once this has been achieved, they can each locally consider the rebuilt values, come to a decision, and then begin a Byzantine agreement process to make sure that they are all in agreement with the proper way to proceed. While this initially seems like a cumbersome step, recent research has lead to great improvements in Byzantine agreement and fault tolerance protocols \cite{castro1999practical,guerraoui2010next,cowling2006hq}. Given the promising results of these research endeavors, Byzantine agreement has been made considerably more feasible and less costly of an operation than was previously the case. The additional complexity from the Byzantine agreement scheme itself is dependent on the choice of approach though we have proposed a few options an in depth analysis of the complexity of these protocols is somewhat outside the scope of our present work. However, similar to our use, this type of agreement mechanism is increasing in appearance due to some of the discoveries of performance improvements, most notably perhaps in cryptocurrencies \cite{garay2015bitcoin,decker2016bitcoin}.
\subsection{Secure Distributed Firewall Protocol 2}

\begin{algorithm}[!htbp]
	\caption{\newline firewallEval$(\langle Gateway, addr\rangle, \langle P_1,\perp\rangle, \dots \langle P_m,\perp\rangle,)\newline\rightarrow  (\langle Gateway, \pi\rangle, \langle P_1, \perp\rangle\dots \langle P_m, \perp \rangle) $\label{alg:fireprod}}
	\DontPrintSemicolon
	\SetKwInput{KwData}{Input}
	\SetKwInput{KwResult}{Output}
	\KwData{IP address (addr) or other information of interest, parties should be in possession of shares of the firewall represented as a Bloom filter, this is a vector $[\phi_j]_N^{P_i}$ for $j=0\dots r-1, i=1\dots m$. Public info: $N$ the modulus of the secret sharing scheme}
	\KwResult{$\pi$ is revealed by the Gateway }
	Gateway
	\begin{enumerate}[label=(\alph*)]
		\item $addr\gets$ IP address (or other info of interest) extracted from packet
		\item Send $addr$ to $P_i$, for $i=1\dots m$		
	\end{enumerate} 
	\nl$P_i$ for $i=1\dots m$ 
	\begin{enumerate}[label=(\alph*)]
		\item $[\pi]_N^{P_i}=\prod_j[\phi_j]_N^{P_i}: j\in \{hash_q(addr)\mod \beta, \forall q \in {1\dots\kappa}\}$
		\item Send $[\pi]_N^{P_i}$ to Gateway
	\end{enumerate} 
	\nl Gateway 
	\begin{enumerate}[label=(\alph*)]
		\item $reveal(\pi) $ for all valid combinations of $t$ out of $m$ shares 
		\item if all ${m\choose t}$ reveals do not agree flag malicious behavior
		\item if $\pi = 1$: block packet
		\item else: forward packet
	\end{enumerate}
\end{algorithm}

\subsubsection{Key steps and correctness analysis}
The steps in this version of the protocol are all exactly identical with one small, through very significant difference. Instead of a summation a product is called for in step 2. This results in large changes for both the complexity and the amount of information to be hidden. Now the revealed values will be either 0 or 1 exclusively. The revealed result will be one iff every indexed element involved in the product is 1. This is again directly analogous to the result of checking for set membership using a Bloom filter directly. This additional information hidden is the contents of the filter itself, which may be advantageous for some use-cases. 

\subsubsection{Complexity analysis for Protocol 2}
The initialization is the same as previously, the first transmitted message is the same, and the conclusion of the protocol is the same with all parties sending their share of the result to the Gateway. Thus the complexity is the same for these steps, requiring $m(32+\log_2N)$ bits. On top of this we add some additional communications and operations due to the fact that products cannot be evaluated without interaction among the parties. Assuming that Shamir's secret sharing scheme is the scheme in use and that $m=2t-1$, the unbounded fan-in product drives up the cost. 

The additional cost for this operation, as we gave previously in Section \ref{sec:fanmult} is in generating $2(\kappa+1)$ random values, evaluating $3\kappa+2$ binary input products, and cooperating in $2\kappa+1$ reveal operations in 5 rounds. Generating a shared random value requires all $m$ parties to send shares of a locally generated random value to the other $m-1$ parties. Thus this operation requires $m(m-1)\log_2N$ bits each. The complexity of a standard binary secure multiplication in Shamir's scheme is the same. Revealing a shared value is achieved simply by all parties exchanging their shares and locally rebuilding the secret, which is again the same complexity. Thus the communication complexity for this protocol version is $(7\kappa+5)(m(m-1)\log_2N)+m(32+\log_2N)$ in 6 rounds.  

\subsubsection{Proofs of security for a semi-honest adversary}\label{sec:sec2}
With respect to security, the arguments of the previous protocol hold, and we additionally strengthen the ability to hide the contents of the filter itself. We do so by considerably lengthening the expected time to the filter's contents becoming known through the alteration from summation to multiplication of the Bloom filter indices mentioned previously. 

Consider the previous protocol, Given a large number of queries, information may accrue which will allow the contents and structure of the Bloom filter to be reconstructed. Revealing the Bloom filter does not necessarily imply a leak of information with respect to the firewall we are using it to represent. Nevertheless, our second protocol prevents even the filter's structure from being revealed. This information may be extracted over time with the previous protocol version by saving a record of IP addresses and reconstructed results at the Gateway. Imagine a packet arrives with an IP address that hashes to 4,5, and 6. If the rebuilt summation of these locations equals 2, then obviously 2 of the previously mentioned indices. Imagine a second packet arrives with an address which hashes to 1, 5, and 6. In this case too let us assume that the revealed value is 2. Finally a third packet arrives with an address which hashes to 1, 4, and 7. The revealed summation result here we shall take as 0 for this example. Given these three instances, we can deduce with certainty that the Bloom filter has these contents $[?_0,0_1,?_2,?_3,0_4,1_5,1_6,0_7]$ for the respective indices given in the subscripts. 

Again this may be tolerable in some instances and it may be preferable to hide this in others. In the previous example, if our second proposed protocol were followed, the result for every case would be 0. Therefore the vector would still consist of only uncertainties rather than having some of the values known. Information may still be gleaned from this second protocol by similar means, but the rate of its accrual is much slower since only combinations which return a 1 will yield any sure foundations for deduction. 

\subsection{Performance Evaluation}
\subsubsection{Initialization}
\begin{figure}[]
	\centering
	\begin{minipage}{.47\columnwidth}
		\centering
		\includegraphics[width=\linewidth]{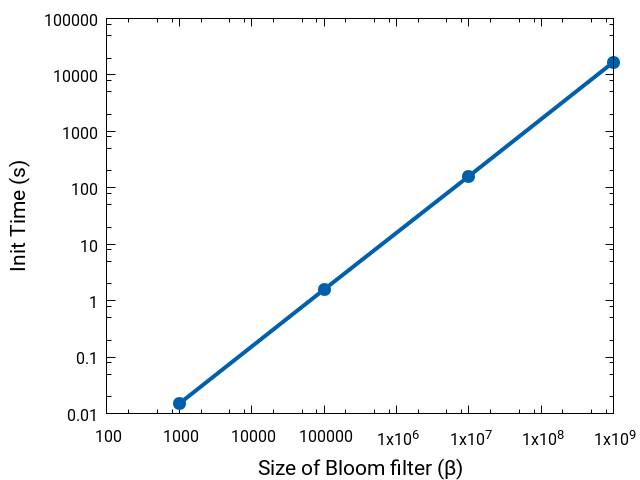}
		\caption{Runtime of system initialization as a function of $\beta$ with $m$ fixed\label{gra:initPerfbeta}}
	\end{minipage}\hspace{.5cm}\begin{minipage}{.47\columnwidth}
		\centering
		\includegraphics[width= \linewidth]{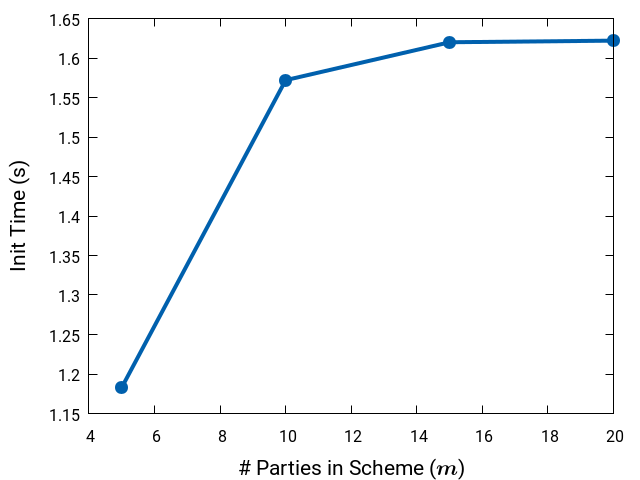}
		\caption{Runtime of system initialization as a function of $m$ with $\beta$ fixed \label{gra:initPerfm}}
	\end{minipage}\vspace{-.5cm}
\end{figure}

The runtime of this process is plotted as a function of the number of locations in the Bloom filter $\beta$ and $m$ the number of parties in the secret sharing scheme. We have covered a very large range in $\beta$ from the necessary width for about 100 entries in the firewall to nearly  1\% of the IPv4 address space with a small error rate selected of .01. This is diagrammed in Figures \ref{gra:initPerfbeta}, add \ref{gra:initPerfm}. Note the use of a log scale on the $\beta$ axis and for the runtime axis and color gradient. When graphed with respect to $\beta$, $m$ is fixed at 10, and when graphed with respect to $m$, $\beta$ is fixed at For firewalls containing a very large number of entries, this process can get somewhat lengthy, but need only be done once. Additionally, this performance test has been done under Python and without parallelization. In practice, implementation with c or c++ and taking advantage of parallelization would both yield dramatic performance improvements for large firewall. Online performance is much better, even for large firewalls, as we will presently show.

%

\subsubsection{Firewall Evaluation}
\begin{figure}[]
	\centering
	\vspace{.5cm}
	\begin{minipage}{.47\columnwidth}
		\centering
		\includegraphics[width=\linewidth]{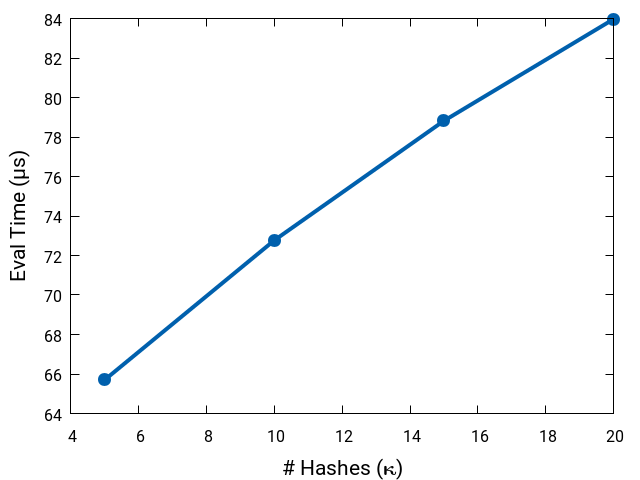}
		\caption{Runtime of Algorithm \ref{alg:firesum} as a function of $\kappa$ with $m$ fixed\label{gra:sumPerfbeta}}
	\end{minipage}\hspace{.5cm}\begin{minipage}{.47\columnwidth}
		\centering
		\includegraphics[width= \linewidth]{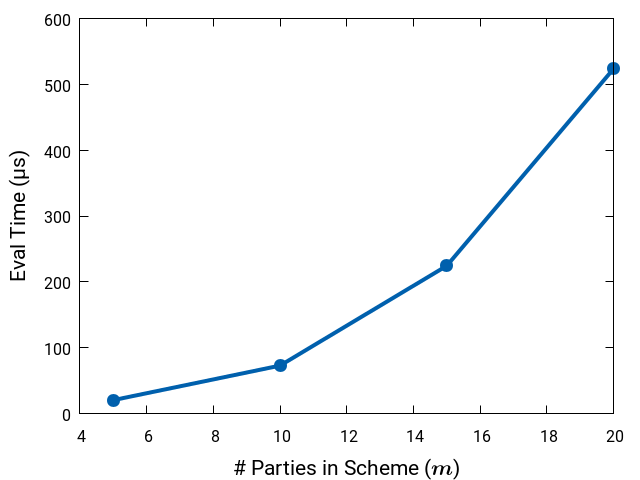}
		\caption{Runtime of Algorithm \ref{alg:firesum} as a function of $m$ with $\kappa$ fixed\label{gra:sumPerfm}}
	\end{minipage}\vspace{-.5cm}
\end{figure}

A graphical depiction of the simulated runtime of this protocol as a function of $\kappa$ and $m$ is given in Figures \ref{gra:sumPerfbeta}, and \ref{gra:sumPerfm}. As can be seen by inspection, Even a maximal $\kappa$ of 20 and $m=20$ the evaluation of the firewall function via the Bloom filter is still well under a millisecond. Additionally, a $\kappa$ of 20 will allow the error rate of the Bloom filter to be less than .00001 with appropriate $\beta$ and $\eta$ values as discussed in Section \ref{sec:bloom}.


%

\subsection{Additional Operations}
We also propose in this section two additional possibilities which may be of interest. The first extends the capabilities of our approach making additions to the firewall represented in the context of the Bloom filter extensible over time. We allow the insertion of new elements following the initialization and sharing of the filter via a simple and efficient means. Secondly, we propose a variation of our protocol which allows for the decision to be made collectively rather than being made solely in the Gateway regarding the rejection or forwarding of packets. This will of course drive up complexity, but in return it also provides potentially more incentive for honesty as malicious activities become even harder to hide. 
\subsubsection{Adding elements to the Firewall}
In many situations, subsequent to the establishment of security measures, new threats may be identified, and there will therefore be a strong desire to include such threats in a blacklist. If the list structure were static, this could pose a serious problem since a new structure would have to be constructed and shared from scratch. Thus we propose a simple means by which elements may be added to the Bloom filter representing the firewall rules without disturbing elements already encoded in that structure. Assuming some kind of strict access control structure is in place, on each server participating in the secret sharing scheme, the following computation must be executed for the new address ($addr$) to be inserted in the filter. This operation is invoked by a trusted system administrator.

\[ \begin{aligned}
\ [\phi_j]_N^{P_i}= &[\phi_j]_N^{P_i}+1-[\phi_j]_N^{P_i} \\ &: j\in \{hash_q(addr)\mod \beta\ \forall q \in {1\dots\kappa\}}
\end{aligned} \]

Care must be taken in this operation since the shared values in the Bloom filter should be exclusively 0 or 1 we cannot allow this manipulation to introduce shares of elements aside from those two. Thus we cannot simply increment the shares for each location resulting from the modular hash of the address in question. In the above expression each location, iff it is previously 0 gets incremented to 1. Additionally, if it is previously 1, it remains unchanged, as desired.  
\cbstart
\subsubsection{More complex rules}
The rules which have been considered up to this point have been entirely based on some combination of source and destination IP address and perhaps port information. It is possible however for our proposed system to handles rules of greater complexity. Individual Internet protocols may be considered or various packet flags. Any information contained within a packet may be the focus of a rule under our system. For separate criteria however one must either greatly expand the magnitude of the Bloom filers in use to cover the additional cases represented by these additional criteria, or, as we argue more strongly for, multiple filters are constructed, evaluated in parallel, and the results of the evaluation composed to form the final result of the evaluation concerning whether the packet should be permitted to pass the firewall, or whether it should be dropped. Given the case of presented in Algorithm \ref{alg:fireprod}. Multiple filters could be constructed based on focus, e.g. one for source IP address, one for destination IP address, one for source port, one for destination port, one for protocol type, etc. Once the individual results of each filter have been calculated, but before these results are returned to the server, the product of all the individual filter results should be calculated with the appropriate primitive of the secret sharing scheme being used. Thus iff the result of every filter is 1, the resulting product will still be one. If for any individual filter, or set of filters, the result is 0, the product will obviously be 0. This step has the added benefit of hiding from the server or any other party, which of the criteria caused the packet to be dropped. While this does add additional complexity to the scheme, it is still able to be done in constant rounds, and is well within the realms of practical realization.  
\cbend

\section{Conclusion}\label{sec:conclusion}
Our contribution in this paper amounts to a suite of protocols for a secure, distributed, and oblivious, firewall evaluation scheme. We have achieved this new and innovative paradigm through the use of secret sharing schemes as well as Bloom filters to represent the firewall and affect the evaluation of a packet's membership in the rules of the firewall. We have proposed multiple means for implementing this functionality with various options to increase efficiency or security, including fault tolerance, and security against malicious adversaries attempts to disrupt the system. Furthermore, we have shown that our proposed approach need not be static, but can be dynamically updated if necessary to handle new threats or other security issues. All of this we achieve with reasonably low overhead depending on design choices and security concerns. Overall this method presents a novel set of ideas which are for the first time presented in the literature to achieve firewall distribution and obfuscation in this way.

\Urlmuskip=0mu plus 1mu
\bibliographystyle{IEEEtran}

\end{document}